\documentclass[twocolumn]{aastex7}

\usepackage{amsmath}
\usepackage{float}

\begin{document}

\title{Submillimeter observations of the white dwarf pulsar AR Sco}

\correspondingauthor{Paul E. Barrett}

\author[0000-0002-8456-1424]{Paul E. Barrett}
\affiliation{The George Washington University \\
725 21st Street NW \\
Washington, DC, 20052, USA}
\email{pebarrett@gwu.edu}

\author[0000-0003-0685-3621]{Mark A.~Gurwell}
\affiliation{Center for Astrophysics $|$ Harvard \& Smithsonian\\
60 Garden Street \\
Cambridge, MA 02138, USA}
\email{mgurwell@cfa.harvard.edu}

\begin{abstract}

AR~Scorpii, the so called white dwarf pulsar, contains a rapidly rotating magnetic white dwarf (WD; P$_{spin}$ = 117.0564 s) interacting with a cool, red dwarf (RD) companion in a 3.56 hour orbit. It is a strong radio source with an inverted spectral index between 1--200~GHz that is indicative of synchrotron emission. This paper presents the first submillimeter observations of AR~Scorpii using the Submillimeter Array, helping to fill a critical gap in the spectral energy distribution between 10--600~GHz. The average flux densities at 220 and 345~GHz are 124 and 86~mJy, respectively. The lower than expected flux density at 345~GHz suggests a break in the synchrotron emission at about 200~GHz. A periodogram analysis of the 220~GHz observations shows a modulation with an amplitude of $\approx 6$\% at a period of 58.26~s or at twice the spin frequency of the white dwarf. This modulation is the first direct detection of the WD spin period at radio frequencies and implies that the synchrotron emission arises near the WD and and not from an interaction with the photosphere of the RD. A fit to the spectral energy distribution shows that the synchrotron emission arises from a small, low density region with a magnetic field of 43~MG at a distance of 0.6 orbital radii from the WD. This result implies that AR Scorpii contains a weakly magnetic WD ($\sim 15$~MG) and not a strongly magnetic WD ($\sim 500$~MG) as previously asserted.

\end{abstract}




\section{Introduction} \label{sec:introduction}

The discovery of pulsar-like behaviour in two close white dwarf-red dwarf binaries AR Scorpii (Porb = 3.56 hr; henceforth AR Sco) and J191213.72-441045.1 (Porb = 4.03 hr) has firmly demonstrated that WDs can exhibit many of the same characteristics as neutron star pulsars. The non-thermal emission from the spinning white dwarf (WD) in AR Sco is highly pulsed (up to 90\% pulse fraction) and spans over six orders of magnitude in frequency from the radio to the ultraviolet \citep{2016Natur.537..374M}. The binary shows no flickering or flaring at optical wavelengths indicating that the binary is detached and there is no accretion onto the WD, hence the nomenclature of WD pulsar. The pulsations are seen at the 118 s beat period between the 117 s spin period and the orbital period of the binary. Perhaps the most intriguing aspect of AR Sco is that the bulk of the luminosity of the system is a result of the spin-down power of the WD, which is measured to be dP/dt = –($4.82 \pm 0.18$)~10$^{-17}$~Hz~s$^{-1}$ \citep{2020MNRAS.496.4849G}. By equating the spin-down loss rate to the magnetic dipole loss rate, the WD is inferred to have a very strong magnetic field ($\sim 800$ MG). To explain the WD’s strong magnetic field and fast spin rate, it is proposed that the initial magnetic field of the WD was weak ($\sim 1$~MG), allowing the WD to be spun up by an accretion disk during an earlier evolutionary phase of high accretion. As the WD cooled, its core crystallized causing a convective layer to form, resulting in a rotation-driven dynamo that increased the magnetic field by a factor of ~800 \citep{2021NatAs...5..648S}. The core crystallization scenario has significant ramifications on our understanding of cataclysmic variable (CV) evolution by showing that a strong magnetic field ($\sim 100$~MG) can arise late in the evolution of a CV and not necessarily during the initial common envelope phase as is commonly believed. On the other hand, a weak magnetic field implies another mechanism is responsible for the spin-down loss rate other than magnetic dipole radiation. Therefore, understanding the synchrotron emission from AR Sco can be used to test this scenario by providing an estimate of the magnetic field strength of the WD.

The overall spectral energy distribution of AR Sco was shown by \citet{2016Natur.537..374M} to be dominated by two nonthermal power-law ($S_{\nu} \propto \nu^{\alpha}$) components, although the thermal contribution from the M5 red dwarf (RD) companion is also evident. From radio to infrared frequencies ($\nu \leq 10^9$--$10^{13}$~Hz) the power law slope is $\alpha \approx 1.3$, typical of synchrotron emission, while for the higher optical to X-ray frequencies ($\nu \geq $few$ \times 10^{14}$~Hz), $\alpha \approx -0.2$ \citep{2016Natur.537..374M, 2016ApJ...831L..10G}. However, the highest frequency (X-ray) constraint was quite loose, based only a short Swift Target of Opportunity (ToO) observation. More recent observations of AR Sco have extended the wavelength range, time coverage, and sensitivity. X-ray observations with XMM-Newton, reported by \citet{2017ApJ...851..143T}, shows both orbital and beat variations, and an X-ray spectrum characterized by a hot multi-temperature thermal plasma ($kT$ = 1--8~keV, \citealt{2017ApJ...851..143T}). AR Sco was observed at radio frequencies by \citet{2017A&A...601L...7M}, \citet{2019AAS...23334809S} and \citet{2018A&A...611A..66S} using  the Australian Long Baseline Array (LBA), the Very Long Baseline Array (VLBA) and the Karl G. Jansky Very Large Array (VLA), respectively. The source was unresolved in the LBA and VLBA observations. All observations reported strong modulations of the radio flux on the orbital and beat periods. Long cadence optical photometry, utilizing Kepler data, has been reported in \citet{2017ApJ...845L...7L}.

This paper presents the first observations of AR Sco at 1.4 and 0.87~mm using the Submillimeter Array (SMA). Sections 2 and 3 present the observations and results, respectively. In section 4, we argue that the radio emission between 1--10~GHz is due to electron cyclotron maser emission from near the RD and between 10--600~GHz to fast-cooling synchrotron emission. We present our conclusions based on the results of our analysis in section 5.

\section{Observations} \label{sec:observations}

AR Sco was observed twice using the SMA. The first observation was on 2022 April 9, with the local oscillator (LO) frequency near 220~GHz (1.4 mm), and the second observation was 2022 May 16 with the LO frequency near 345~GHz (0.87 mm) (see Table \ref{tab:obs}). For each observation, dual polarization receivers were utilized with 12~GHz of bandwidth in each of two sidebands per each polarization, for a total of 48 GHz of processed bandwidth. A per scan integration time of 14.84 s was implemented, resulting in 9.9 min of AR Sco integration time between visits to nearby radio-loud AGN J1625-254 and J1700-261, that were used as complex gain calibrators. Each calibrator was observed for 1.24 minutes. Each observation resulted in roughly 45 to 52 minutes total integration time on AR Sco, spanning a little over 1 hour. Both observations were conducted with six antennas operating in the SMA 'extended' configuration, with maximum baselines to 220~m, providing spatial resolutions of $\approx 1.3$" and 0.85" at 220 and 345~GHz, respectively. The flux density scale was measured using observations of Ceres, MWC349a, and Ganymede obtained at different times during operations each night. Standard data reduction procedures were used to produce fully calibrated visibility data of AR Sco. Because the apparent angular scale of the AR Sco binary ($< 1$ milli-arcsec) is much smaller than the spatial resolution of the instrument ($\sim1$ arcsec), flux densities were measured by fitting a simple point source ($\delta$-function) to the calibrated visibilities using the software Visfit.jl \citep{2025inprep}.

\begin{table}
    \centering
    \caption{Log of SMA observations of AR Sco.}
    \label{tab:obs}
    \begin{tabular}{lcccc}
    \hline \hline
    Mean      &    Date    &    Start   &    Stop    & Exposure \\
    Frequency &            &    Time    &    Time    &          \\
    (GHz)     &            &    (UTC)   &    (UTC)   &    (s)   \\
    \hline
    219.609   & 2022-04-09 & 14:56:46.5 & 16:01:05.2 &  2686 \\
    345.292   & 2022-05-16 & 10:24:47.8 & 11:34:33.1 &  3116 \\
    \hline
    \end{tabular}
\end{table}

\section{Results} \label{sec:results}

\subsection{Spectral Measurements} \label{subsec:fluxes}



AR Sco was observed twice using the SMA for 2686 and 3116~s at 220 and 345~GHz, respectively. The average flux density at 220~GHz was $124 \pm 2$~mJy with a range that varied between 77 and 167~mJy. The average flux density at 345~GHz was $86 \pm 11$~mJy. The spectral index $\alpha = 0.358 \pm 0.015$ in the 5--9~GHz band \citep{2018A&A...611A..66S} and appears to increase at higher frequencies. Our observations support this conclusion by measuring an $\alpha = 0.515 \pm 0.015$ between 22--220~GHz and $\alpha = -0.3 \pm 0.1$ between 220--345~GHz. Therefore, the SMA observations indicate either a break or turn-over in the spectrum at about 220~GHz. If there was no change in spectral slope above 220~GHz, then the flux density at 345~GHz would be $\approx 150$~mJy or nearly twice the observed value. The implications of the spectral break are discussed in greater detail in subsection \ref{subsec:SED}.

\subsection{Time Series Analysis} \label{subsec:timeseries}

\begin{figure}
    \label{fig:SMA_timeseries}
    \centering
    \includegraphics[width=\columnwidth]{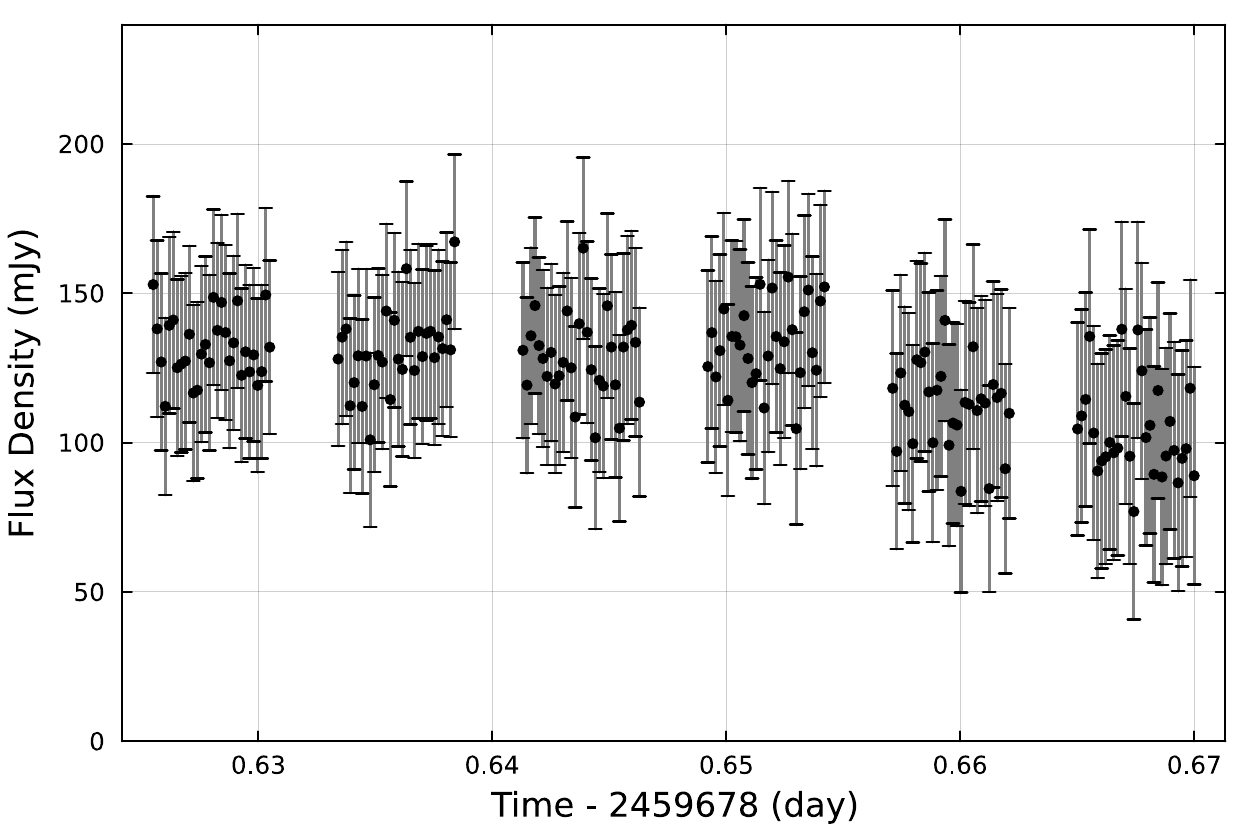}
    \caption{Photometry of AR Sco at 220~GHz.}
\end{figure}

\begin{figure}
    \label{fig:SMA_brett1}
    \centering
    \includegraphics[width=\columnwidth]{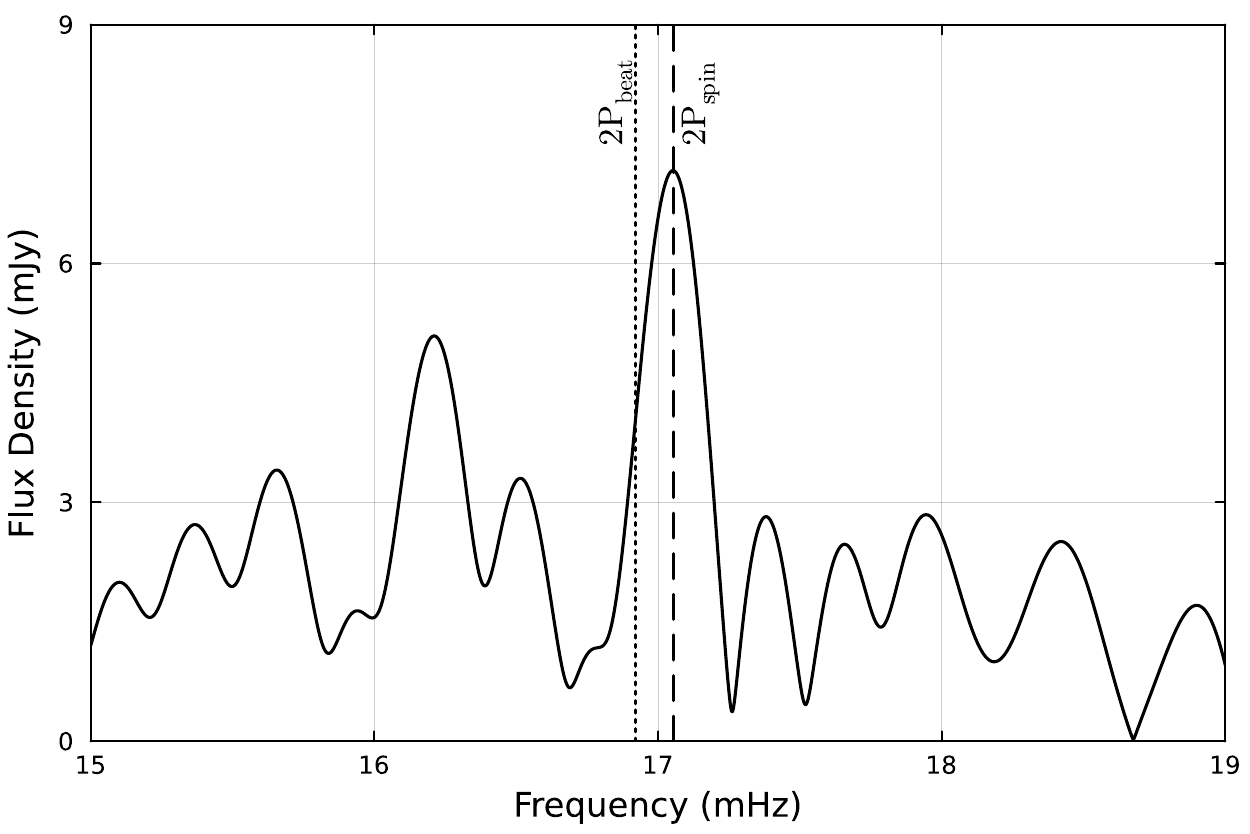}
    \caption{Periodogram of AR Sco photometry at 220~GHz. The signal at $17.053 \pm 0.023$~mHz with an amplitude of $\approx 6$\% is identified as the second harmonic of the spin period (117.12 s). The dashed vertical line is the frequency of twice the spin period and the dotted vertical line is that of twice the beat period. The difference is $>5\sigma$. Most amplitudes are $<3$ mJy.}
\end{figure}

\begin{figure}
    \label{fig:SMA_spin_phase}
    \centering
    \includegraphics[width=\columnwidth]{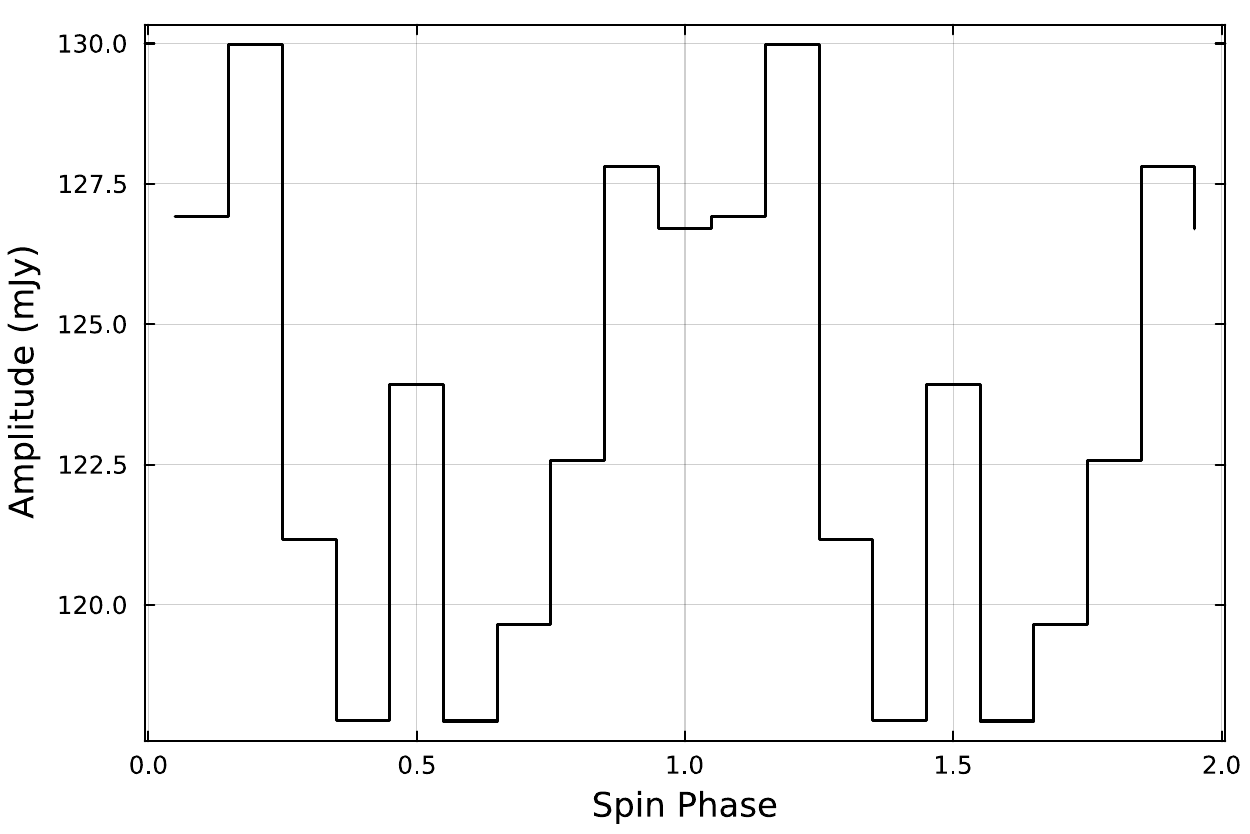}
    \caption{Spin folded light curve of AR Sco at 220~GHz.}
\end{figure}

The 220~GHz observation has sufficient sensitivity at an integration time of 14.84 s to generate a light curve (see Fig. \ref{fig:SMA_timeseries}). A Bretthorst periodogram \citep{Bretthorst1988Bayesian} analysis of the light curve is shown in Figure \ref{fig:SMA_brett1}. There is one clear signal at 17.053(23)~mHz (58.641(79)~s), where the values in parentheses are the error in the last digits. \cite{2022MNRAS.516.5052P} give the spin frequency of the white dwarf and its second harmonic to be 8.5382182845(30) and 17.0764365689(59)~mHz, respectively. We therefore attribute the SMA signal to be a modulation at twice the spin frequency of the WD and not at twice the beat frequency. The SMA signal differs from the beat frequency by $> 5 \sigma$ as is shown in Figure \ref{fig:SMA_brett1}. This signal is the first direct detection of the spin period of the WD in AR Sco at radio frequencies. Previous observations at other frequencies, particularly in the optical, have only detected the beat period and have inferred the spin period from the orbital and beat periods. The amplitude of the SMA signal is 7.1~mJy or 5.7\% of the average flux density. The spin folded light curve is shown in Figure \ref{fig:SMA_spin_phase}.

\section{Discussion}

\subsection{Spectral Energy Distribution} \label{subsec:SED}

\citet{2017ApJ...851..143T, 2018ApJ...853..106T} propose that most of the emission from the radio to X-rays ($10^{9}$--$10^{18}$~Hz) can be explained by a single synchrotron emission component with an electron power law index of 2.5--3. This component dominates the emission over more than five orders of magnitude in frequency from 1~GHz to greater than 100~THz (see e.g., \citealt{2017ApJ...851..143T}, Figure 10). This model of the non-thermal emission is a rough approximation to the observational data, particularly in the frequency range of 1--200~GHz. Its first spectral break and peak emission occurs in the far infrared at a few~THz, while the second spectral break occurs in the $\gamma$-ray band at a few tens of MeV. The model is a reasonable fit to the spectrum above the peak emission (i.e., from the infrared to X-rays), but it overestimates the observed spectrum below the peak by an order of magnitude or more because the spectral index $\alpha = 1/3$ is too flat for the observed spectrum. Although the VLA observations of AR Sco by \cite{2018A&A...611A..66S} find an average spectral index $\alpha = 0.358 \pm 0.015$ in the 5--9~GHz range, this is in a narrow frequency range that, as noted below, involves two emission components. Our 220~GHz observations along with an unpublished 22~GHz VLA observations (Barrett, private communication) fills in the gap between 10--600~GHz as shown in Figure \ref{fig:AR_Sco_SED} and listed in Table \ref{tab:fluxes} of Appendix \ref{appendix:fluxes}. These observations provide critical information about the spectral energy distribution (SED) in this frequency range. These results show that the 10--220~GHz spectral index is steeper ($\alpha = 0.515 \pm 0.015$) than that measured in the 4--10~GHz range. The SMA observations also show that the first spectral break is at $\approx 200$~GHz, because the flux density at 345~GHz without a break would be $\approx$150~mJy/bm, which is about twice the observed value of 86~mJy/bm. Because of these observational constraints, we calculate a new SED for AR Sco containing six emission components. Three components are due to thermal emission from the WD, the RD, and a cool circumbinary dust cloud having temperatures of 9750, 3100, and 70~K, respectively. The RD and the dust emission provide most of the emission in the near and far infrared parts of the spectrum. The other three components are due to non-thermal emission caused by the rapidly rotating WD.

\citet{2018A&A...611A..66S} note that there appears to be two non-thermal components in the radio part of the spectrum with a transition region at about 10~GHz. Their VLA observations show the polarized emission to be modulated at the orbital period. The linear polarization is typically $<$1\%, while the circular polarization varies between 0 and $-30$\%. They attribute this polarized emission to non-relativistic cyclotron emission. The problem with cyclotron emission is that it produces narrow emission lines at harmonics of the gyro-frequency that decrease rapidly with increasing harmonic number resulting in most of the emission occurring from a few narrow emission lines of the first few harmonics \citep[see e.g.,][]{1984ApJ...278..298B}. Instead, we suggest that electron cyclotron maser emission (ECME) is more consistent with the highly circularly polarized radio emission below 10~GHz. Above 10 GHz, partially absorbed synchrotron emission dominates, because it is unpolarized and modulated at the spin and beat periods. The third non-thermal component is the X-ray emission from the synchrotron emission region and the surface of the WD, because the X-ray emission at the spin period is seen during part of the orbital period, while the emission at the beat period is seen throughout the orbital period.

\subsection{The Electron Cyclotron Maser Emission Component}

Electron cyclotron maser emission is a common radio emission process in magnetic cataclysmic variables \citep{2017AJ....154..252B, 2020AdSpR..66.1226B}. The physical conditions required for its emission are very different from that of synchrotron radiation. First, the plasma frequency of the medium must be less than the gyro-frequency. At GHz frequencies, this constrains the electron density to be $\lesssim 10^{12}$ cm$^{-3}$. Second, the magnetic field of the plasma must be $\sim 4$ kG for the emission to occur at frequencies of $\geq 1$~GHz. Neither of these conditions satisfied those found in the synchrotron emission region. These plasma conditions are similar to those of the lower corona of the red dwarf. It seems unlikely that ECME would occur in a low density region between the two stars, because of the rapid rotation of the WD. If the ECME is from the red dwarf, then its magnetic moment is $\sim 2 \times 10^{34}$ G cm$^3$, is typical for rapidly rotating stars, because the magnetic dynamo saturates for rotation rates of $<10$ days \citep{1999A&A...346..922K, 2008ApJ...676.1262B}.

\subsection{Time Averaged Synchrotron Emission Model} \label{subsec:Model}

We interpret the change in the spectral slope of the synchrotron emission at $\approx 220$~GHz to be a break in the spectrum and attribute it to partial absorption. Such breaks are common features of synchrotron shock models from $\gamma$-ray bursts (GRBs). There are various external shock models to describe the spectral and temporal behaviour of GRBs depending on whether the dominant synchrotron emission arises from a forward or reverse, thick or thin-shell shock expanding into a constant or decreasing density medium, i.e., into the interstellar medium or a stellar wind, respectively. They are defined by three frequencies: the self absorption frequency, $\nu_a$; the cooling frequency, $\nu_c$; and the frequency of the minimum injected energy of the electrons, $\nu_m$. $p$ is the power-law index of the electron distribution. The three frequencies determine the location of the spectral breaks. In most cases the self-absorption frequency $\nu_a << \nu_c, \nu_m$ and this appears to be the case for AR Sco. This constraint narrows the various spectral regimes from six to two; namely, the fast cooling regime ($\nu_c < \nu_m$) and the slow cooling regime ($\nu_m < \nu_c$). Unlike GRBs whose shocks evolve over many days, the synchrotron emission from AR Sco is believed to be the result of a continuous injection of relativistic electrons caused by magnetic reconnection or magnetic dipole radiation. The temporal behaviour of these shock models is therefore not important, only their spectral characteristics. Therefore, we will use the simplest model, the thin-shell forward shock model, to fit the spectral energy distribution of AR Sco from the radio to optical frequencies. See Appendix \ref{appendix:power} for details.

\subsection{The Synchrotron Component} \label{subsec:constraints}

Both synchrotron models have seven adjustable parameters: those specifying the spectral breaks, $\nu_a$, $\nu_c$, and $\nu_m$; those specifying the electron distribution, $n_e$, and $p$; and those specifying the emission region, $B$, $R$, and L, where $R$ is its size and $L$ is the thickness of the absorbing medium. The bulk motion of the medium $\Gamma_2$ is assume to be non-cosmological and is therefore set to 1. The relativistic motion of the electrons $\gamma_{c,m}$ is assumed to be $\sim 10$. Previous and current radio observations of AR Sco place two constraints on the two emission models. They are the unpolarized emission at 22 GHz and the size of the emission region. The unpolarized emission at 22~GHz (Barrett, private communication) implies that the emission at this frequency is partially absorbed. The VLBA observation of AR Sco at X-band is unresolved at a spatial resolution of $\approx 0.5$~mas \citep{2019AAS...23334809S}. At the distance of AR Sco, this result constrains the size of the radio emission region $\lesssim 10^{12}$ cm.

\begin{figure*}
    \centering
    \includegraphics[width=6.5in]{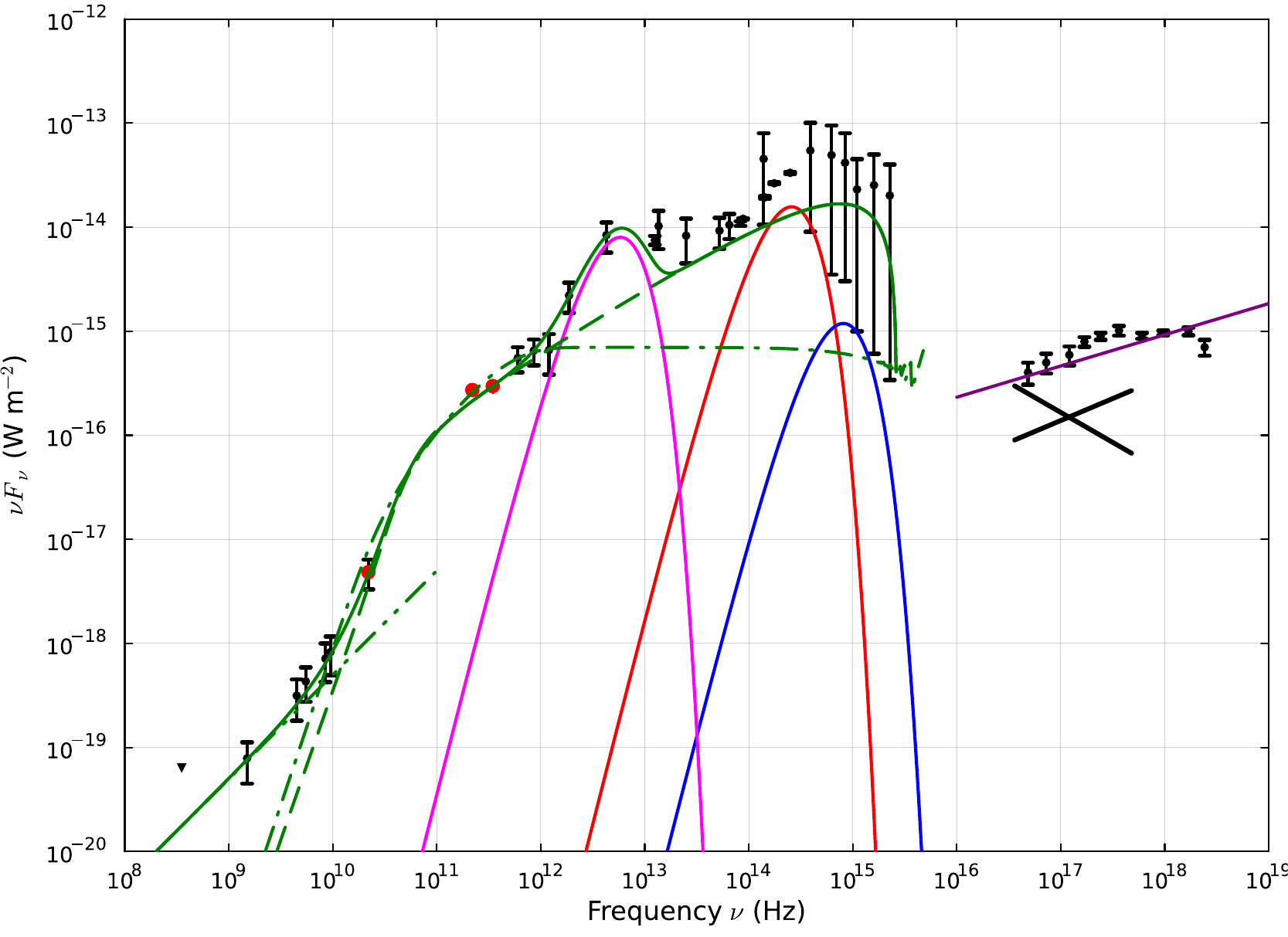}
    \caption{Spectral energy distribution (SED) of AR Sco. The black points and lines are data from \citet{2016Natur.537..374M} and \citet{2018ApJ...853..106T}. The three red points are the 22 GHz (VLA) plus 220 and~345 GHz (SMA) observations. The magenta, red, and blue curves are respectively the blackbody emissions from the cool circumbinary dust cloud (70 K), the red dwarf companion (3100 K), and the white dwarf primary (9750 K). The dashed and dash-dotted green lines are respectively the fast and slow cooling models for the synchrotron emission ($> 10$ GHz). The dash-dot-dotted green line is the electron cyclotron maser emission ($< 10$ GHz) and the purple line is the power-law X-ray emission. The solid green line is the sum of the various emission components.}
    \label{fig:AR_Sco_SED}
\end{figure*}

An initial fit to the synchrotron spectrum is done by eye. The seven parameters are adjusted to obtain a good fit to the spectral slope below 220 GHz, while maximizing the emission at higher frequencies. The analysis begins by fitting the spectral slope below 220 GHz because this spectral region determines the frequencies for $\nu_a$ and $\nu_{c,m}$, where $\nu_a$ is dependent on the parameters $n_e$, $B$, $R$, and $L$. Values of $B$, $n_e$, $R$, and $L$ of 20~G, $1.5 \times 10^9$~cm$^{-3}$, $3 \times 10^7$~cm, and $3 \times 10^5$~cm for the fast cooling model and 20 G, $2 \times 10^{10}$~cm$^{-3}$, $1.2 \times 10^7$~cm, and $1 \times 10^6$~cm for the slow cooling model sets $\nu_a$ at $< 1$ GHz. Next, $\nu_c$, $\nu_m$, and $p$ are fit to the measurements above 200 GHz. A good fit gives values of $3 \times 10^{10}$~Hz, $5 \times 10^{15}$~Hz, and 1.6 for the fast cooling model and $1 \times 10^{15}$~Hz, $3 \times 10^{11}$~Hz, and 3 for the slow cooling model, respectively. Markov Chain Monte Carlo (MCMC) sampling is then performed to find the optimal values and standard deviations of $\nu_c$, $\gamma_e$, $n_e$, $B$, $R$, and $L$ for the fast cooling model using the fit-by-eye values as starting points. For the sampling to converge, $\nu_a$, $\nu_m$, and $p$ are held fixed, because they have little impact on the spectrum between 1 GHz and 1 THz. The results of the sampling suggest a higher magnetic field strength, similar sizes for the emission and absorption regions, and lower values for the cooling frequency and electron density. The result of the spectral fit is summarized in Table \ref{tab:models}, where the quoted errors are the Markov chain standard error (MCSE). The best fast and slow cooling models are shown in Figure \ref{fig:AR_Sco_SED}. Although both models fit the slope well in the radio part of the spectrum, the fast-cooling model appears to be a better overall fit to the SED in the infrared and optical part of the spectrum. It also partially fills in the region between the cool dust and red dwarf blackbody emission.

The MCMC result suggests a magnetic field of the emission region of $\approx 43$ G and a small emission size of $\sim 10^7$~cm or $\approx 0.1$\% of the orbital radius. The small size of the emission region provides little constraint on its location between the two stars. A few models place it close to the RD star, i.e., at a distance of $\sim 1.2 \times 10^{11}$~cm from the WD \citep{2018ApJ...853..106T}. The problem with this scenario is that the low frequency radio emission (1--10~GHz) is probably due to ECME from the RD, which implies a surface magnetic field of $\sim 4$~kG. The synchrotron emission therefore cannot be emitted near the RD surface. Other models place it midway between the two stars \citep{2017NatAs...1E..29B}. Assuming a 4 kG dipolar magnetic field for the RD and the synchrotron emission region at the mid-point between the two stars, the strength of the RD magnetic field at this location is $\approx 114$~G, which implies a WD dipolar magnetic field of $> 185$~MG. The magnetic field estimate from the MCMC result indicates that the synchrotron emission region is even further from the RD at a distance of $\approx 1.4$ orbital radii or at a distance of 0.6 orbital radii from the WD. At this distance the polar field strength of the WD is $\approx 15$~MG, similar to the field strength of the intermediate polar (IP) subclass of magnetic CVs. The field strength is consistent with an upper limit of 100~MG from an analysis of combined interpulse far-ultraviolet spectra from the Hubble Space Telescope \citep{2021ApJ...908..195G}. This result does not support the assertion that AR Sco has undergone a phase of WD crystallization. Instead, AR Sco appears to be an IP whose accretion has temporarily ceased.

\begin{table}
\caption{Synchrotron Model Parameters}
\label{tab:models}
\begin{tabular}{lccc}
\hline \hline
 Parameter         &  Fast-cooling  &  Slow-cooling  \\
 \hline
 $\nu_a$ (Hz)      & $                 < 10^9     $     & $         < 10^{9} $ \\
 $\nu_c$ (Hz)      & $(6.4   \pm 0.3)   \times 10^9  $  & $1   \times 10^{15}$ \\
 $\nu_m$ (Hz)      & $5                 \times 10^{15}$ & $2   \times 10^{11}$ \\
 $\gamma_e$        & $10.2   \pm 0.1$                   &  10                  \\
 p                 &  1.1                               &  3                   \\
 n$_e$ ($cm^{-3}$) & $(5.34  \pm 0.07)  \times 10^8 $   & $2   \times 10^{10}$ \\
 B (G)             & $42.7   \pm 0.2$                   &  20                  \\
 R (cm)            & $(3.786 \pm 0.005) \times 10^7 $   & $1.2 \times 10^{7}$  \\
 L (cm)            & $(4.61  \pm 0.05)  \times 10^5 $   & $1   \times 10^{6}$  \\
 \hline
 \end{tabular}
\end{table}

\section{Conclusions} \label{sec:conclusions}

The submillimeter SMA and high frequency (22~GHz) VLA observations of AR Sco help fill the gap between the low frequency VLA radio and far-infrared Herschel observations. The SMA observations provide critical constraints on the emission process and location. The detection of twice the WD spin frequency at 220~GHz shows that the submillimeter emission arises from an interaction between the dipolar magnetic field of the WD and the magnetic field of the RD companion, and not from an interaction with the photosphere of the RD. Otherwise, the emission would be modulated at the beat period or its harmonic. Given the fast timescale ($\sim$ seconds) required to accelerate electrons to relativistic energies, either magnetic reconnection or magnetic dipolar radiation are likely to be the primary injection mechanism for the synchrotron emission. A fit to the SED of AR Sco from the radio to optical frequencies gives a magnetic field strength of $\approx 43$~G for the synchrotron emission region. This result implies that the WD has a weak magnetic field ($\sim 15$~MG), which is contrary to the strong magnetic field ($\sim 500$~MG) based on the spin-down power of the WD. Therefore, AR Sco does not appear to have undergone a phase of WD crystallization. Additional high time resolution ($\sim 1$~s) and high precision ($\sim 10$~mJy) millimeter and submillimeter observations spanning an entire orbital period combined with a more accurate model that simulates the motion of the magnetic poles throughout the WD rotation would help to confirm the conclusion that the WD in AR Sco has a weak magnetic field.


\begin{acknowledgments}
We thank the Submillimeter Array for providing Director's Discretionary Time (DDT) for these exploratory observations. We also thank AvdH for valuable discussions about synchrotron emission models and AB for discussions of the VLBA observations of AR Sco. The Submillimeter Array is a joint project between the Smithsonian Astrophysical Observatory and the Academia Sinica Institute of Astronomy and Astrophysics and is funded by the Smithsonian Institution and the Academia Sinica. We recognize that Maunakea is a culturally important site for the indigenous Hawaiian people; we are privileged to study the cosmos from its summit.
\end{acknowledgments}

%

\vspace{5mm}
\facilities{SMA(radio), VLA(radio)}


\software{Astronomical Image Processing System \citep{1985daa..conf..195W}, Visfit.jl (Barrett, P. 2025, in preparation)}

\appendix
\section{Synchrotron Power and Absorption Coefficients}
\label{appendix:power}

The luminosity for a one dimensional emitting region with absorption is:
\begin{equation}
L_{\nu} = \frac{4\pi R^3}{3 \alpha_{\nu} L} P_{\nu} (1 - exp(-\alpha_{\nu} L)), 
\end{equation}
where the observed synchrotron powers and absorption coefficients are approximately:
\begin{widetext}
    I. $\nu_a < \nu_c < \nu_m$: \\

    \begin{equation}\label{equ:pow_fast}
    \begin{split}
    P_{\nu}& = 3 \left(\frac{3}{2}\right)^{\frac{1}{3}} \frac{n_e q_e^3 B \Gamma_2}{m_e c^2} \frac{1}{2}
             \left[ 1 - \frac{p-1}{p}\left(\frac{\nu_m}{\nu_c}\right)^{-\frac{1}{2}} \right]^{-1} \\
             & \quad \times \Bigg\{ \left(\frac{\nu}{\nu_c}\right)^{-\frac{1}{2}}
             \left[\Gamma\left(\frac{5}{6},\frac{\nu}{\nu_m}\right) -
             \Gamma\left(\frac{5}{6},\frac{\nu}{\nu_c}\right)\right] \\
             & \quad + \left(\frac{\nu_m}{\nu_c}\right)^{-\frac{1}{2}}\left(\frac{\nu}{\nu_m}\right)^{-\frac{p}{2}} \left[ \Gamma\left(\frac{p}{2}+\frac{1}{3}\right)-\Gamma\left(\frac{p}{2}+\frac{1}{3}, \frac{\nu}{\nu_m}\right) \right] \Bigg\},
    \end{split}
    \end{equation}

    \begin{equation}\label{equ:abs_fast}
    \begin{split}
    \alpha_{\nu}& = 3 \left(\frac{3}{2}\right)^{\frac{1}{3}} \frac{n_e q_e^3 B \Gamma_2}{8 \pi m_e^2 c^2 \gamma_c \nu^2} \frac{1}{2}
            \left[ 1 - \frac{p-1}{p}\left(\frac{\nu_m}{\nu_c}\right)^{-\frac{1}{2}} \right]^{-1} \\
            & \quad \times \Bigg\{ 4 \left(\frac{\nu}{\nu_c}\right)^{-1} \left[\Gamma\left(\frac{4}{3},\frac{\nu}{\nu_m}\right) - \Gamma\left(\frac{4}{3},\frac{\nu}{\nu_c}\right)\right] \\
            & \quad + (p+3) \left(\frac{\nu_m}{\nu_c}\right)^{-1} \left(\frac{\nu}{\nu_m}\right)^{-\frac{p+1}{2}} \left[ \Gamma\left(\frac{p}{2}+\frac{5}{6}\right)-\Gamma\left(\frac{p}{2}+\frac{5}{6}, \frac{\nu}{\nu_m}\right) \right] \Bigg\},
    \end{split}
    \end{equation}

II. $\nu_a < \nu_m < \nu_c$:

    \begin{equation}\label{equ:pow_slow}
    \begin{split}
    P_{\nu}& = 3 \left(\frac{3}{2}\right)^{\frac{1}{3}} \frac{n_e q_e^3 B \Gamma_2}{m_e c^2} \frac{p-1}{2}
            \left[ 1 - \frac{1}{p}\left(\frac{\nu_c}{\nu_m}\right)^{-\frac{p-1}{2}} \right]^{-1} \\
           & \quad \times \Bigg\{ \left(\frac{\nu}{\nu_m}\right)^{-\frac{p-1}{2}} \left[\Gamma\left(\frac{p}{2}-
           \frac{1}{6},\frac{\nu}{\nu_c}\right) - \Gamma\left(\frac{p}{2}-\frac{1}{6},\frac{\nu}{\nu_m}\right)\right] \\
           & \quad + \left(\frac{\nu_c}{\nu_m}\right)^{-\frac{p-1}{2}}\left(\frac{\nu}{\nu_c}\right)^{-\frac{p}{2}} \left[ \Gamma\left(\frac{p}{2}+\frac{1}{3}\right)-\Gamma\left(\frac{p}{2}+\frac{1}{3}, \frac{\nu}{\nu_c}\right) \right] \Bigg\},
    \end{split}
    \end{equation}
    
    \begin{equation}\label{equ:abs_slow}
    \begin{split}
    \alpha_{\nu}& = 3 \left(\frac{3}{2}\right)^{\frac{1}{3}} \frac{n_e q_e^3 B \Gamma_2}{8 \pi m_e^2 c^2 \gamma_m \nu^2} \frac{p-1}{2}
            \left[1 - \frac{1}{p}\left(\frac{\nu_c}{\nu_m}\right)^{-\frac{p-1}{2}} \right]^{-1} \\
           & \quad \times \Bigg\{ (p+2) \left(\frac{\nu}{\nu_m}\right)^{-\frac{p}{2}} \left[\Gamma\left(\frac{p}{2} +
           \frac{1}{3},\frac{\nu}{\nu_c}\right) - \Gamma\left(\frac{p}{2}+\frac{1}{3},\frac{\nu}{\nu_m}\right)\right] \\
           & \quad + (p+3) \left(\frac{\nu_c}{\nu_m}\right)^{-\frac{p}{2}} \left(\frac{\nu}{\nu_c}\right)^{-\frac{p+1}{2}} \left[ \Gamma\left(\frac{p}{2}+\frac{5}{6}\right)-\Gamma\left(\frac{p}{2}+\frac{5}{6}, \frac{\nu}{\nu_c}\right) \right] \Bigg\}
    \end{split}
    \end{equation}
    
\end{widetext}
In the previous equations, $n_e$, $q_e$, and $m_e$ are the electron density, charge, and mass, respectively; $c$ is the speed of light; $B$ is the magnetic field strength; and $\Gamma_2$ is the cosmological Lorentz factor of the source. $\Gamma(x)$ and $\Gamma(x, y)$ in equations (\ref{equ:pow_fast}--\ref{equ:abs_slow}) are the complete and incomplete gamma functions, respectively \citep[see, e.g.][]{1986rpa..book.....R, 2013NewAR..57..141G, 2021MNRAS.506.4275B}.

\section{Sources and Fluxes}
\label{appendix:fluxes}

Table \ref{tab:fluxes} is a list of archival and observed fluxes of AR Sco that span more than ten orders of magnitude in frequency. The fourth and eighth (Orbit) columns specify observations that cover a full or partial orbit. The 22, 220, and 345~GHz VLA and SMA flux densities are new measurements. All other measurements are from archival data from \citet{2016Natur.537..374M} and \citet{2018ApJ...853..106T}.

\begin{table*}[t]
    \centering
    \caption{Data Table: Archival and Observed Fluxes}
    \label{tab:fluxes}
    \begin{tabular}{lrrrlrrr}
    \hline \hline
    Source  & Frequency &    Flux Density     &  Orbit  &  Source & Frequency &     Flux Density          &  Orbit  \\
    \hline
    WISH     &  352 MHz &          $<18$ mJy  & partial & VLT      &  139 THz &  $7.55 - 57.6$ mJy        &   full  \\
    VLA      &  1.5 GHz &   $3.0 -  7.5$ mJy  &    full & 2MASS    &  143 THz & $13.5 \pm 0.3$ mJy        & partial \\
    VLA      &  4.5 GHz &   $4.0 - 10.0$ mJy  &    full & 2MASS    &  176 THz & $15.0 \pm 0.3$ mJy        & partial \\
    VLA      &  5.5 GHz &   $5.0 - 10.7$ mJy  &    full & 2MASS    &  250 THz & $13.3 \pm 0.3$ mJy        & partial \\
    VLA      &  8.5 GHz &   $5.0 - 11.8$ mJy  &    full & WHT      &  391 THz &   $2.3 - 25.6$ mJy        &   full  \\
    VLA      &  9.5 GHz &   $5.2 - 12.2$ mJy  &    full & WHT      &  625 THz &  $0.56 - 15.2$ mJy        &   full  \\
    VLA      & 22.0 GHz &    $22 \pm  1$ mJy  &    full & WHT      &  843 THz &  $0.36 -  9.5$ mJy        &   full  \\
    SMA      &  220 GHz &   $124 \pm 10$ mJy  & partial & WHT      & 1.10 PHz & $0.091 -  4.1$ mJy        &   full  \\
    SMA      &  345 GHz &    $86 \pm 11$ mJy  & partial & HST      & 1.60 PHz & $0.038 - 3.13$ mJy        &   full  \\
    Herschel &  600 GHz &    $92 \pm 25$ mJy  & partial & HST      & 2.27 PHz & $0.015 - 1.76$ mJy        &   full  \\
    Herschel &  857 GHz &    $76 \pm 21$ mJy  & partial & XMM      & 48.4 PHz & $0.83 \pm 0.2$ $\mu$Jy    &   full  \\
    Herschel & 1.20 THz &    $55 \pm 23$ mJy  & partial & XMM      & 72.5 PHz & $0.69 \pm 0.15$ $\mu$Jy   &   full  \\
    Herschel & 1.87 THz &   $118 \pm 38$ mJy  & partial & XMM      &  121 PHz & $0.49 \pm 0.1$ $\mu$Jy    &   full  \\
    Herschel & 4.28 THz &   $196 \pm 63$ mJy  & partial & XMM      &  169 PHz & $0.47 \pm 0.05$ $\mu$Jy   &   full  \\
    Spitzer  & 12.5 THz &   $59.6 \pm 6$ mJy  & partial & XMM      &  242 PHz & $0.37 \pm 0.03$ $\mu$Jy   &   full  \\
    WISE     & 13.6 THz & $45.2 - 105.4$ mJy  & partial & XMM      &  363 PHz & $0.28 \pm 0.03$ $\mu$Jy   &   full  \\
    WISE     & 25.0 THz &  $18.0 - 48.3$ mJy  & partial & XMM      &  605 PHz & $0.15 \pm 0.01$ $\mu$Jy   &   full  \\
    Spitzer  & 52.3 THz &  $11.9 - 23.5$ mJy  & partial & XMM      &  967 PHz & $0.1 \pm 0.005$ $\mu$Jy   &   full  \\
    WISE     & 65.1 THz &  $11.8 - 20.5$ mJy  & partial & XMM      & 1.69 EHz & $0.059 \pm 0.005$ $\mu$Jy &   full  \\
    Spitzer  & 83.3 THz & $13.0 \pm 0.7$ mJy  & partial & XMM      & 2.42 EHz & $0.029 \pm 0.005$ $\mu$Jy &   full  \\
    WISE     & 88.2 THz &  $13.5 - 13.8$ mJy  & partial &          &          &                           &         \\
    \hline
    XMM      &   36 PHz &       0.25 $\mu$Jy  &         &  \dots   &  480 PHz &      0.056 $\mu$Jy        &   full \\
    XMM      &   36 PHz &       0.83 $\mu$Jy  &         &  \dots   &  480 PHz &      0.014 $\mu$Jy        &   full \\
    \hline
    \end{tabular}
\end{table*}




\bibliography{SMA_AR_Sco}{}
\bibliographystyle{aasjournal}



\end{document}